\documentclass[10pt,conference]{IEEEtran}
\usepackage[T1]{fontenc}
\usepackage[utf8]{inputenc}
\usepackage{cite}
\usepackage{amsmath,amssymb,amsfonts}
\usepackage{booktabs}
\usepackage{array}
\usepackage{multirow}
\usepackage{graphicx}
\usepackage{xcolor}
\usepackage{upquote}
\usepackage{url}
\usepackage{microtype}
\usepackage{enumitem}
\usepackage{balance}

\setlist{nosep, leftmargin=*, topsep=2pt}
\begin{document}

\title{Specification-Grounded Adaptive Triage and Formal Invariant Exclusion Synthesis for Autonomous Silicon Coverage Closure}

\author{\IEEEauthorblockN{Veni Mohan}
\IEEEauthorblockA{Google LLC, Mountain View, CA, USA\\
Email: venimohan@google.com}}

\maketitle

\begin{abstract}
Closing the final 5\%--15\% of code and functional coverage (``the long tail of verification'') remains one of the primary schedule bottlenecks in modern silicon tapeouts. Recent attempts to automate coverage closure using Large Language Models (LLMs) feed raw uncovered RTL lines or branches directly into iterative prompt loops. While effective on small toy blocks, such \emph{flat, unclustered LLM loops} fail on industrial subsystems due to three fundamental limitations: (1) \textbf{Token \& Context Explosion}, as thousands of correlated coverage holes stemming from a single unexercised operating mode are queried independently; (2) \textbf{The Oracle \& Illegal-Stimulus Problem}, where unconstrained LLMs synthesize stimulus that violates cross-signal microarchitectural invariants, yielding ``dirty'' coverage or scoreboard failures; and (3) \textbf{Unsafe Exclusion Hallucination}, where structurally unreachable dead code (e.g., compile-time parameter masking, address-width padding, mutually exclusive FSM states, or subsystem tie-offs) stalls stimulus generation or induces hallucinated waivers.

We present \textbf{ARGUS} (\emph{Autonomous Reachability \& Guided Unified Stimulus}), a specification-grounded, dual-regime coverage closure and formal exclusion synthesis engine. First, ARGUS ingests microarchitectural specifications and SystemVerilog Assertions (SVAs) to construct a formal \textbf{Block Contract} ($\mathcal{C}$). Second, rather than treating uncovered points flatly, ARGUS introduces an adaptive \textbf{Macro-Clustering vs.\ Micro-Surgery Triage Engine} parameterized by a per-module hole-density threshold ($\tau$). Modules exceeding $\tau$ undergo \emph{Macro-Clustering}, collapsing thousands of correlated line, branch, condition, and toggle holes into a compact set of root-cause operational mode, FSM, and backpressure clusters. Modules at or below $\tau$ transition to \emph{Micro-Surgery}, performing Abstract Syntax Tree (AST) and loop-domain reachability analysis to partition remaining holes into reachable corner-case stimulus targets vs.\ mathematically unreachable structural invariants. Finally, ARGUS pairs directed test synthesis with bit-exact golden scoreboards and automatically emits tool-native \texttt{.el} exclusion files backed by formal invariant proofs.

We evaluate ARGUS across \textbf{five core hardware benchmarks} spanning three open-source \textbf{LowRISC OpenTitan} security/peripheral IPs (\texttt{uart\_core}, \texttt{i2c\_core}, and \texttt{aes\_cipher\_core}) and two commercial advanced-node \textbf{AI Accelerator Tensor Layout Converter (TLC)} datapaths (\texttt{Layout-3D-Subsys} and \texttt{Layout-MultiPipe-Top}) totaling \textbf{53 SystemVerilog modules and 376,127 coverage points}, and further demonstrate enterprise scalability on a \textbf{2,626-module industrial AI accelerator interconnect subsystem} (\texttt{Industrial-ICN-2626}, \texttt{Industrial-Fetch-Top}, and \texttt{Industrial-DRAM-Arb}). Across a 4-way quantitative ablation study and comparisons against \textbf{LLM-for-DV and commercial formal baselines}, ARGUS triages 2,626 modules in \textbf{0.265 seconds}, reduces LLM token consumption by \textbf{27.2$\times$ to $>$180$\times$}, achieves \textbf{zero dirty merges} via contract-constrained 3-step sanity gating, and reaches \textbf{99.32\%--100.0\% effective coverage closure} with \textbf{100.0\% formal exclusion soundness (0 false exclusions)} across all \textbf{21,071+ proven exclusions}.
\end{abstract}

\section{Introduction}

In modern industrial Application-Specific Integrated Circuit (ASIC) development---spanning energy-efficient embedded systems~\cite{mohan2018dac}, security-critical SoC peripherals, and hyperscale AI accelerators---design verification (DV) consumes over 60\% of total engineering effort. While constrained-random verification (CRV) in Universal Verification Methodology (UVM) environments rapidly achieves 80\%--85\% code and functional coverage, closing the remaining 15\%--20\% to meet strict tapeout signoff criteria (typically $\ge 98\%$ Line, Branch, Condition, and Toggle coverage) requires weeks to months of manual engineering effort~\cite{ioannides2012todaes,sv2023}.

Recently, academic and industrial researchers have proposed using Large Language Models (LLMs) to automate RTL code, assertion, and test generation from coverage reports~\cite{orenes2023llm,wan2026chattest,liu2024chipnemo,orenes2021autosva,tsai2024rtlfixer,zhang2023llm4dv,thakur2024verigen,kande2024tifs}. However, deploying existing flat LLM-in-the-loop frameworks on real silicon exposes three severe failure modes:

\begin{enumerate}
    \item \textbf{Correlated Hole Redundancy \& Token Explosion:} In production RTL, a single unexercised operating mode (e.g., an unexercised target-mode clock-stretching FSM in OpenTitan I2C, an unexercised AES-192/256 decryption key-expansion path in OpenTitan AES, or an inactive 32-bit vector mode in an AI accelerator datapath) leaves hundreds of lines, thousands of multi-term boolean conditions, and tens of thousands of register bits uncovered. A flat agent that queries an LLM per hole burns millions of tokens rediscovering the same root cause.
    \item \textbf{The Oracle Problem \& Illegal Mode Combinations:} Hardware blocks enforce strict cross-parameter and temporal invariants---for example, requiring multi-rail sparse handshake encodings (\texttt{SP2V\_HIGH = 3'b011}, \texttt{SP2V\_LOW = 3'b100}) in OpenTitan \texttt{aes\_cipher\_core}, exact SCL/SDA setup/hold timing windows in \texttt{i2c\_fsm}, or $C_{\max} \le C_{\text{pack}}$ during tensor pooling. Prompting an LLM solely with local RTL snippets produces stimulus that triggers coverage counters while violating architectural contracts (``dirty coverage'') or failing golden scoreboard checks.
    \item \textbf{Unreachable Structural Invariants vs.\ Reachable Corners:} In parameterized RTL, 3.5\% to 35\% of raw coverage points are structurally unreachable by construction---such as unmasked PRNG/S-Box mask buses (\texttt{gen\_no\_masks} tied to \texttt{'0} when \texttt{SecMasking = 0} in OpenTitan AES), 32-bit bus-timeout upper bits (\texttt{scl\_high\_cnt[31:11]} in OpenTitan I2C), upper address bits of a 16-bit bus addressing a 128-row SRAM (\texttt{addr[15:7]}), or mutually exclusive FSM state conditions (\texttt{fsm\_idle \&\& fsm\_done\_q}). Without formal reachability separation, an agent either loops infinitely trying to hit dead code or unsafely waives reachable corner cases.
\end{enumerate}

To overcome these challenges, we architected and deployed \textbf{ARGUS}, an autonomous coverage closure and formal exclusion synthesis framework. This paper makes the following contributions:
\begin{itemize}
    \item \textbf{Specification-Grounded Block Contracts ($\mathcal{C}$):} We formulate an automated extraction pass (\texttt{SpecExtractor}) that mines operational modes, legal parameter domains, and interface invariants from microarchitectural specifications and embedded RTL SVAs to constrain downstream stimulus synthesis.
    \item \textbf{Adaptive Dual-Regime Triage ($\tau$-Switching):} We introduce \texttt{HoleClassifier}, a hierarchical triage algorithm that dynamically switches between \textbf{Macro-Clustering} (when module holes $> \tau$) to collapse correlated mode/stall/FSM holes into single root-cause stimulus recipes, and \textbf{AST Micro-Surgery} (when module holes $\le \tau$) to separate reachable micro-corners from unreachable invariants.
    \item \textbf{Oracle-Paired Stimulus \& Formal \texttt{.el} Exclusion Synthesis:} We pair every synthesized test with bit-exact reference scoreboard checks (\texttt{StimulusSteerer}) and prove structural unreachability across 12 mathematical invariant classes to emit production-ready VCS/URG \texttt{.el} exclusion files (\texttt{ExclusionGenerator}).
    \item \textbf{Cross-Domain Evaluation Across 5 Subsystems \& 4-Way Ablation:} We evaluate ARGUS across \textbf{three open-source LowRISC OpenTitan IPs} (\texttt{uart\_core}, \texttt{i2c\_core}, \texttt{aes\_cipher\_core}) and \textbf{two production AI accelerator Tensor Layout Converter subsystems} (\texttt{Layout-3D-Subsys} and \texttt{Layout-MultiPipe-Top}), demonstrating a \textbf{27.2$\times$--95.7$\times$ token reduction}, \textbf{zero dirty merges} via 3-step sanity gating, and \textbf{100\% formal exclusion soundness} across \textbf{376,127 coverage points}.
\end{itemize}

\section{Preliminaries \& Problem Formulation}

Let an RTL design under test (DUT) consist of a set of hierarchical modules $\mathcal{M} = \{m_1, m_2, \dots, m_K\}$. A coverage database $\mathcal{D}$ produced by simulation (e.g., Verilator \texttt{coverage.dat}~\cite{verilator} or Synopsys VCS \texttt{.vdb}~\cite{synopsys_urg}) defines a universe of coverage points:
\begin{equation}
\mathcal{P} = \big\{ p_i = (m_i, \ell_i, t_i, e_i, c_i) \big\}_{i=1}^{|\mathcal{P}|}
\end{equation}
Here, $\ell_i \in \mathbb{N}$ is the source line number, $t_i \in \{\text{line}, \text{branch}, \text{cond}, \text{toggle}, \text{fsm}\}$ is the metric type, $e_i$ is the boolean guard expression or signal bit identifier, and $c_i \in \mathbb{N}_0$ is the execution hit count. The uncovered hole set at iteration $k$ is $\mathcal{U}^{(k)} = \{p_i \in \mathcal{P} \mid c_i^{(k)} = 0\}$.

\subsection{The Partitioning \& Closure Objective}

Not all uncovered points $p \in \mathcal{U}^{(0)}$ can be exercised by legal input stimulus. Let $\mathcal{I}_{\text{spec}}$ denote the set of all input trace sequences permitted by the block's microarchitectural specification and SVA assumptions (\texttt{assume property}). The uncovered set $\mathcal{U}^{(0)}$ partitions into two disjoint subsets:
\begin{equation}
\mathcal{U}^{(0)} = \mathcal{U}_{\text{reach}} \;\dot{\cup}\; \mathcal{U}_{\text{unreach}}
\end{equation}
where:
\begin{itemize}
    \item \textbf{Reachable Holes ($\mathcal{U}_{\text{reach}}$):} Points $p \in \mathcal{U}^{(0)}$ for which there exists at least one legal stimulus trace $\sigma \in \mathcal{I}_{\text{spec}}$ that drives $c_p(\sigma) > 0$ while satisfying all golden scoreboard post-conditions.
    \item \textbf{Structurally Unreachable Invariants ($\mathcal{U}_{\text{unreach}}$):} Points $p \in \mathcal{U}^{(0)}$ for which $\forall \sigma \in \mathcal{I}_{\text{spec}}, c_p(\sigma) = 0$ by mathematical construction (e.g., compile-time parameter masking, 2-state/sparse enum defaults, shift-register framing invariants, loop domain bounds, one-hot state mutual exclusion, or static tie-offs).
\end{itemize}

The goal of autonomous coverage closure is to simultaneously synthesize a minimal test suite $\mathcal{T} = \{\sigma_1, \dots, \sigma_N\} \subset \mathcal{I}_{\text{spec}}$ that covers $\mathcal{U}_{\text{reach}}$ and a formal exclusion set $\mathcal{E}$ that covers $\mathcal{U}_{\text{unreach}}$ subject to \textbf{strict soundness} ($\mathcal{E} \cap \mathcal{U}_{\text{reach}} = \emptyset$, i.e., zero false exclusions).

\section{ARGUS Architecture \& Methodology}

ARGUS is architected as a six-stage modular pipeline comprising \texttt{CoverageParser}, \texttt{SpecExtractor}, \texttt{HoleClassifier}, \texttt{StimulusSteerer}, \texttt{ExclusionGenerator}, and \texttt{ReportGenerator}.

\subsection{Unified Coverage Ingestion (\texttt{CoverageParser})}

To operate seamlessly across fast unit-level C++/Verilator testbenches and full-chip Synopsys VCS UVM regressions, \texttt{CoverageParser} normalizes heterogeneous coverage databases into a unified in-memory representation (\texttt{CoverageDatabase}). For Verilator \texttt{SystemC::Coverage-3} databases, \texttt{CoverageParser} decodes \texttt{\textbackslash x01}/\texttt{\textbackslash x02} key-value metadata frames, extracts metric pages (\texttt{v\_line}, \texttt{v\_branch}, \texttt{v\_expr}, \texttt{v\_toggle}), strips template parameterization suffixes, and separates RTL logic modules from hard SRAM macro wrappers. For Synopsys VCS \texttt{.vdb} and URG hierarchies, it ingests per-instance Line, Condition, Branch, FSM, and Toggle tables alongside \texttt{.el} exclusion checksums.

\subsection{Specification \& SVA Contract Extraction (\texttt{SpecExtractor})}

To prevent the synthesis of illegal mode combinations, \texttt{SpecExtractor} parses natural-language microarchitectural specification documents (\texttt{.md}), register package headers (\texttt{uart\_reg\_pkg.sv}, \texttt{i2c\_reg\_pkg.sv}, \texttt{aes\_pkg.sv}), golden layout tables, and RTL source files (\texttt{.sv}/\texttt{.svp})~\cite{opentitan,vasudevan2010goldmine,gao2023hyde} to construct a formal \texttt{BlockContract}:
\begin{equation}
\mathcal{C} = \big(\mathcal{O}_{\text{modes}},\; \mathcal{R}_{\text{params}},\; \Phi_{\text{inv}}\big)
\end{equation}
\begin{enumerate}
    \item \textbf{Operational Modes ($\mathcal{O}_{\text{modes}}$):} Extracts mutually exclusive and orthogonal mode configurations---e.g., OpenTitan UART \texttt{parity\_en}/\allowbreak\texttt{rxblvl}, OpenTitan I2C \texttt{host\_enable}/\allowbreak\texttt{target\_enable}/\allowbreak\texttt{stretch\_en\_addr\_tx}, OpenTitan AES \texttt{CIPH\_FWD}/\allowbreak\texttt{CIPH\_INV}, \texttt{AES\_128}/\allowbreak\texttt{AES\_192}/\allowbreak\texttt{AES\_256}, \texttt{dec\_key\_gen}, and TLC \texttt{dest\_is\_dram} $\in \{0,1\}$, data types $\{8\text{b}, 16\text{b}, 32\text{b}\}$, \texttt{vec\_half\_mode}, \texttt{pool\_mode\_i}, and \texttt{ch\_pack\_mode}.
    \item \textbf{Parameter Domains ($\mathcal{R}_{\text{params}}$):} Extracts legal dimension ranges, sparse encodings (\texttt{SP2V\_HIGH = 3'b011}, \texttt{SP2V\_LOW = 3'b100}), FIFO depths ($32$ in UART, $64$ in I2C FMT/RX/TX/ACQ), and TLC tensor dimensions $H, W \in \{4, 8, 12, 16\}$ with boundary crop offsets $\{0, 1, 2, 3\}$ and channel depths $C \in \{8n \mid 8 \le C \le 128\}$.
    \item \textbf{SVA \& Interface Invariants ($\Phi_{\text{inv}}$):} Extracts embedded \texttt{assume property} and \texttt{ASSERT} clauses (\texttt{AesCiphOpValid}, \texttt{AesKeyLenValid}, \texttt{a\_pool\_ch\_bound}).
\end{enumerate}

\subsection{Adaptive Dual-Regime Triage (\texttt{HoleClassifier})}

Coverage holes exhibit \textbf{bimodal spatial locality} across the verification lifecycle. Early in verification (or when a feature mode is disabled in regression), a single module $m$ exhibits $|\mathcal{U}_m| \gg \tau$ uncovered points that share a common control-flow ancestor. Conversely, late in verification ($|\mathcal{U}_m| \le \tau$), the remaining holes are isolated corner cases or structural dead code.

\begin{figure}[t]
\centering
\scriptsize
\resizebox{\columnwidth}{!}{%
\begin{tabular}{p{1.05\columnwidth}}
\toprule
\textbf{Algorithm 1:} Adaptive Dual-Regime Hole Triage \& Invariant Classification \\
\midrule
\textbf{Input:} CoverageDatabase $\mathcal{D}$, BlockContract $\mathcal{C}$, RTL AST Sources $\mathcal{S}$, Threshold $\tau$ \\
\textbf{Output:} TriageReport $\mathcal{R} = (\text{MacroClusters}, \text{MicroHoles}, \text{FormalExclusions})$ \\
1: \textbf{for each} module $m \in \mathcal{D}.\text{modules}$ \textbf{do} \\
2: \hspace{1em} $\mathcal{U}_m \leftarrow \{ p \in \mathcal{D}.\text{points}(m) \mid p.\text{covered} == \text{False} \}$ \\
3: \hspace{1em} \textbf{if} $|\mathcal{U}_m| > \tau$ \textbf{then} \hfill \textit{// REGIME 1: MACRO-CLUSTERING} \\
4: \hspace{2em} $\mathcal{B}_{\text{mode}}, \mathcal{B}_{\text{stall}}, \mathcal{B}_{\text{fsm}}, \mathcal{B}_{\text{prox}} \leftarrow \emptyset$ \\
5: \hspace{2em} \textbf{for each} hole $p \in \mathcal{U}_m$ \textbf{do} \\
6: \hspace{3em} $\text{sig\_ctx} \leftarrow \text{Lower}(p.\text{branch\_info} \cup p.\text{hier} \cup \mathcal{S}[m][p.\text{line}])$ \\
7: \hspace{3em} \textbf{if} $\text{MatchesModeKeyword}(\text{sig\_ctx}, \mathcal{C}.\text{modes})$ \textbf{then} $\mathcal{B}_{\text{mode}}[\text{Key}].\text{append}(p)$ \\
8: \hspace{3em} \textbf{else if} $\text{MatchesHandshakeStall}(\text{sig\_ctx})$ \textbf{then} $\mathcal{B}_{\text{stall}}.\text{append}(p)$ \\
9: \hspace{3em} \textbf{else if} $\text{MatchesFSMState}(\text{sig\_ctx})$ \textbf{then} $\mathcal{B}_{\text{fsm}}.\text{append}(p)$ \\
10: \hspace{2.6em} \textbf{else} $\mathcal{B}_{\text{prox}}[\lfloor p.\text{line}/20 \rfloor].\text{append}(p)$ \\
11: \hspace{2em} $\mathcal{R}.\text{MacroClusters}.\text{extend}(\text{EmitMacroRecipes}(\mathcal{B}_{\text{mode}}, \mathcal{B}_{\text{stall}}, \mathcal{B}_{\text{fsm}}, \mathcal{B}_{\text{prox}}, \mathcal{C}))$ \\
12: \hspace{1em} \textbf{else} \hfill \textit{// REGIME 2: MICRO-SURGERY} \\
13: \hspace{2em} \textbf{for each} hole $p \in \mathcal{U}_m$ \textbf{do} \\
14: \hspace{3em} $\text{ast\_node} \leftarrow \text{ParseAST}(\mathcal{S}[m], p.\text{line}, p.\text{branch\_info})$ \\
15: \hspace{3em} \textbf{if} $\text{IsStructurallyUnreachable}(\text{ast\_node}, \mathcal{C}.\text{invariants}, \mathcal{S}[m])$ \textbf{then} \\
16: \hspace{4em} $\text{proof} \leftarrow \text{SynthesizeInvariantProof}(\text{ast\_node}, \mathcal{C})$ \\
17: \hspace{4em} $\mathcal{R}.\text{FormalExclusions}.\text{append}(\text{HoleCluster}(p, \text{FORMAL\_EXCLUSION},$ \\
\hspace{6.2em} $\text{proof}))$ \\
18: \hspace{3em} \textbf{else} \\
19: \hspace{4em} $\text{recipe} \leftarrow \text{SynthesizeCornerStimulus}(\text{ast\_node}, \mathcal{C})$ \\
20: \hspace{4em} $\mathcal{R}.\text{MicroHoles}.\text{append}(\text{HoleCluster}(p, \text{SYNTHESIZE\_MICRO\_TEST},$ \\
\hspace{6.2em} $\text{recipe}))$ \\
21: \textbf{return} $\mathcal{R}$ \\
\bottomrule
\end{tabular}%
}
\end{figure}

As shown in \textbf{Algorithm 1}, when $|\mathcal{U}_m| > \tau$ (default $\tau = 10$), \texttt{HoleClassifier} groups uncovered points into four root-cause Equivalence Buckets: (1) \textbf{Mode-Control Clusters ($\mathcal{B}_{\text{mode}}$)}, (2) \textbf{Handshake \& Backpressure Clusters ($\mathcal{B}_{\text{stall}}$)}, (3) \textbf{FSM \& Protocol Error Clusters ($\mathcal{B}_{\text{fsm}}$)}, and (4) \textbf{Structural Proximity Clusters ($\mathcal{B}_{\text{prox}}$)}. When $|\mathcal{U}_m| \le \tau$ (or during final signoff sweep), \texttt{HoleClassifier} switches to \textbf{Micro-Surgery}, inspecting the exact RTL AST statement and loop bounds to classify each point as either a reachable corner test (\texttt{SYNTHESIZE\_MICRO\_TEST}) or a formally proven unreachable invariant (\texttt{FORMAL\_EXCLUSION}).

\subsection{Oracle-Paired Stimulus Steering \& Formal \texttt{.el} Exclusion Synthesis}

\texttt{StimulusSteerer} binds every directed stimulus configuration to the block's cycle-accurate golden reference model (NIST AES-128/192/256 cipher vectors, I2C Host/Target protocol scoreboard, UART frame monitor, and \texttt{layout\_golden\_model}). For every hole classified into $\mathcal{U}_{\text{unreach}}$, \texttt{ExclusionGenerator} synthesizes a mathematical justification and formats tool-native Synopsys VCS/URG and Verilator \texttt{.el} exclusion files.

\section{Experimental Setup}

To demonstrate that ARGUS generalizes across both \textbf{public open-source peripheral/cryptographic IPs} and \textbf{commercial AI accelerator datapaths}, we evaluate ARGUS across \textbf{five benchmarks} totaling \textbf{53 SystemVerilog modules and 376,127 coverage points}:
\begin{enumerate}
    \item \textbf{Benchmark 1 --- LowRISC OpenTitan \texttt{uart\_core}:} Baud NCO generator, break-detection FSM (\texttt{BRK\_CHK}/\texttt{BRK\_WAIT}), 3-sample majority noise filter, system/line loopback, 24-bit RX timeout counter, TX/RX serializers, dual 32-deep FIFOs (\texttt{prim\_fifo\_sync}), 2-stage synchronizer, and 8 interrupt controllers. \emph{Scale:} 7 SV modules (13 instances), \textbf{1,982 Verilator v5.048 coverage points} (57 Line, 88 Branch, 71 Expr, 1,766 Toggle).
    \item \textbf{Benchmark 2 --- LowRISC OpenTitan \texttt{i2c\_core}:} Dual-mode I2C Host Master and Target Responder FSM (\texttt{i2c\_fsm}, 1,244 lines of RTL with 35 protocol states including \texttt{SetupStart}, \texttt{ClockLowAck}, \texttt{AcquireByte}, \texttt{StretchAddrTx}, \texttt{StretchAddrAcq}, \texttt{StretchTxEmpty}, \texttt{StretchAcqFull}), dual 7-bit target address/mask comparators, SCL/SDA bus-interference detectors, 31-bit stretch and 32-bit host timeout counters, four 64-deep synchronous FIFOs (\texttt{FMT}, \texttt{RX}, \texttt{TX}, \texttt{ACQ}), and 16 interrupt modules. \emph{Scale:} 6 SV modules (24 instances), \textbf{3,906 coverage points} (247 Line, 348 Branch, 247 Expr, 3,064 Toggle).
    \item \textbf{Benchmark 3 --- LowRISC OpenTitan \texttt{aes\_cipher\_core}:} Full AES-128/192/256 forward encryption, inverse decryption, and standalone decryption key generation (\texttt{DecKeyGen}) datapath, dual redundant multi-rail sparse FSMs (\texttt{aes\_cipher\_control\_fsm\_p/n}), on-the-fly key expansion, 20 parallel S-Boxes (16 data + 4 key-expand), \texttt{aes\_shift\_rows}, 4-column $\text{GF}(2^8)$ \texttt{aes\_mix\_columns}, and 18 sparse-encoding fault detectors (\texttt{aes\_sel\_buf\_chk}). \emph{Scale:} 14 SV modules (46 instances), \textbf{19,147 coverage points} (112 Line, 210 Branch, 179 Expr, 18,646 Toggle).
    \item \textbf{Benchmark 4 --- Industrial AI Accelerator \texttt{Layout-3D-Subsys}:} Streams 3D tensor tiles through a credit-managed rolling input buffer, a 3D write-coordinate generator, multi-bank 1R1W ping-pong SRAMs (128 rows/bank), and a single-cycle combinational read decoder/swizzler feeding a shared elastic flop tray across multiple on-chip vector-memory and DRAM packing modes. \emph{Scale:} 16 SV modules comprising \textbf{7,970 RTL coverage points} (313 Line, 376 Branch, 7,281 Toggle) in an industrial neural accelerator datapath.
    \item \textbf{Benchmark 5 --- Industrial AI Accelerator \texttt{Layout-MultiPipe-Top}:} Production multi-lane layout conversion, rotation (\texttt{lane\_rotate}), cropping (\texttt{tile\_crop}), stitching (\texttt{tile\_stitch}), reorder buffering (\texttt{reorder\_buf}), and dual-AXI CSR configuration (\texttt{cfg\_reg\_ctrl}, \texttt{buf\_ctrl}). \emph{Scale:} 10 SV modules across 14 hierarchical instances in a 274MB Synopsys VCS \texttt{.vdb} regression database comprising \textbf{343,122 total coverage points} (14,939 Line, 23,056 Cond, 9,587 Branch, 295,540 Toggle), with \textbf{13,879 uncovered items} requiring triage and closure.
\end{enumerate}

\section{Experimental Results}

\subsection{Multi-Stage Coverage Convergence Across OpenTitan IPs \& Industrial \texttt{Layout-3D-Subsys}}

\textbf{Tables~\ref{tab:uart_i2c}--\ref{tab:aes_tlc}} detail the cumulative Verilator \texttt{-{}-coverage} progression across OpenTitan \texttt{uart\_core}, \texttt{i2c\_core}, \texttt{aes\_cipher\_core}, and industrial \texttt{Layout-3D-Subsys}.

\begin{table}[t]
\centering
\caption{6-Stage Coverage Progression on OpenTitan \texttt{uart\_core} (1,982 Pts) and \texttt{i2c\_core} (3,906 Pts)}
\label{tab:uart_i2c}
\scriptsize
\setlength{\tabcolsep}{2.5pt}
\resizebox{\columnwidth}{!}{%
\begin{tabular}{clccccc}
\toprule
\textbf{Stg} & \textbf{ARGUS-Steered Stimulus Added} & \textbf{Line} & \textbf{Branch} & \textbf{Expr} & \textbf{Toggle} & \textbf{Total} \\
\midrule
\multicolumn{7}{l}{\textit{\textbf{(a) OpenTitan \texttt{uart\_core}} (57 Line, 88 Br, 71 Expr, 1,766 Tgl = 1,982 Pts)}} \\
0 & Seed Smoke (\texttt{sys\_loopback} 8N1 \texttt{0xA5}) & 71.93\% & 72.73\% & 85.92\% & 28.20\% & 33.50\% \\
1 & + Macro 1: Parity, Line Loop, TX Ovrd, Noise & 73.68\% & 80.68\% & 91.55\% & 38.84\% & 43.59\% \\
2 & + Macro 2: 32-Deep FIFO Watermarks \& Ovfl & 91.23\% & 85.23\% & 94.37\% & 57.81\% & 61.30\% \\
3 & + Macro 3: Parity/Frame Err \& Break FSM & 98.25\% & 89.77\% & 94.37\% & 60.87\% & 64.43\% \\
4 & + Macro 4: RX Timeout \& Mid-Sample Glitch & 98.25\% & \textbf{100.0\%} & 94.37\% & 62.46\% & 66.30\% \\
5 & + Micro: 8-Ch Intr Matrix \& $32\times 8$ Sweep & \textbf{98.25\%} & \textbf{100.0\%} & \textbf{98.59\%} & \textbf{96.15\%} & \textbf{96.47\%} \\
5+\texttt{.el} & \textbf{+ Formal \texttt{.el} Invariant Exclusions (70 pts)} & \textbf{100.0\%} & \textbf{100.0\%} & \textbf{100.0\%} & \textbf{100.0\%} & \textbf{100.0\%} \\
\midrule
\multicolumn{7}{l}{\textit{\textbf{(b) OpenTitan \texttt{i2c\_core}} (247 Line, 348 Br, 247 Expr, 3,064 Tgl = 3,906 Pts)}} \\
0 & Seed Smoke (Host 1-Byte Write w/ Start/Stop) & 56.28\% & 44.25\% & 71.26\% & 20.04\% & 27.73\% \\
1 & + Macro 1: Host R/W, Rep Start, PopFmt & 70.04\% & 59.77\% & 74.90\% & 36.10\% & 42.81\% \\
2 & + Macro 2: Target FSM (Addr 0/1, Acq, Tx) & 87.45\% & 79.89\% & 83.40\% & 51.70\% & 58.47\% \\
3 & + Macro 3: Target Clock Stretch FSMs & 90.69\% & 82.18\% & 87.04\% & 52.94\% & 60.09\% \\
4 & + Macro 4: Bus Interference \& Timeouts & 90.69\% & 89.94\% & 88.66\% & 55.55\% & 62.93\% \\
5 & + Micro: 4$\times$64 FIFOs Wrap, StretchAcqFull & \textbf{96.76\%} & \textbf{92.53\%} & \textbf{93.52\%} & \textbf{93.60\%} & \textbf{93.70\%} \\
5+\texttt{.el} & \textbf{+ Formal \texttt{.el} Invariant Exclusions (246 pts)} & \textbf{100.0\%} & \textbf{100.0\%} & \textbf{100.0\%} & \textbf{100.0\%} & \textbf{100.0\%} \\
\bottomrule
\end{tabular}%
}
\end{table}

\begin{table}[t]
\centering
\caption{Coverage Progression on OpenTitan \texttt{aes\_cipher\_core} (19,147 Pts) and Industrial \texttt{Layout-3D-Subsys} (7,970 Pts)}
\label{tab:aes_tlc}
\scriptsize
\setlength{\tabcolsep}{2.5pt}
\resizebox{\columnwidth}{!}{%
\begin{tabular}{clccccc}
\toprule
\textbf{Stg} & \textbf{ARGUS-Steered Stimulus Added} & \textbf{Line} & \textbf{Branch} & \textbf{Expr} & \textbf{Toggle} & \textbf{Total} \\
\midrule
\multicolumn{7}{l}{\textit{\textbf{(a) OpenTitan \texttt{aes\_cipher\_core}} (112 Line, 210 Br, 179 Expr, 18,646 Tgl = 19,147 Pts)}} \\
0 & Seed Smoke (AES-128 Single-Block NIST) & 63.39\% & 36.19\% & 56.98\% & 63.01\% & 62.66\% \\
1 & + Macro 1: AES-128/192/256 Enc/Dec/KeyGen & 84.82\% & 77.14\% & 78.21\% & 67.81\% & 68.10\% \\
2 & + Macro 2: Key/Data-Out PRNG Clearing & 89.29\% & 80.95\% & 81.56\% & 74.94\% & 75.15\% \\
3 & + Macro 3: PRNG Reseed \& FINISH Stall & 91.07\% & 80.95\% & 82.12\% & 75.09\% & 75.31\% \\
4 & + Macro 4: Sparse \texttt{sp2v\_e} Fault \& Alert & 94.64\% & 87.14\% & 84.36\% & 75.35\% & 75.67\% \\
5 & + Micro: 128b State / 256b Key / 20 S-Box & \textbf{94.64\%} & \textbf{87.14\%} & \textbf{84.36\%} & \textbf{78.23\%} & \textbf{78.48\%} \\
5+\texttt{.el} & \textbf{+ Formal \texttt{.el} Unmasked Exclusions (4,120 pts)} & \textbf{100.0\%} & \textbf{100.0\%} & \textbf{100.0\%} & \textbf{100.0\%} & \textbf{100.0\%} \\
\midrule
\multicolumn{7}{l}{\textit{\textbf{(b) Industrial \texttt{Layout-3D-Subsys}} (313 Line, 376 Br, 7,281 Tgl = 7,970 Pts; Clust $\tau=10$)}} \\
0 & Baseline Seed (DRAM 8b $16H\times 16W\times 128C$) & 59.42\% & 56.12\% & --- & 46.16\% & 47.15\% \\
1--3 & + DRAM 8b/16b + VecMem 8b/16b/32b Full/Half Tiers & 87.54\% & 81.12\% & --- & 57.86\% & 60.13\% \\
4--5 & + Multi-Snap ($32\times 32, 256C$), Odd Crops, Pooling & 94.25\% & 88.56\% & --- & 61.34\% & 63.91\% \\
6 & + Backpressure Stalls (50\%, 75\%, Rand) + Micro & \textbf{98.08\%} & \textbf{95.21\%} & --- & \textbf{62.48\%} & \textbf{65.42\%} \\
6+\texttt{.el} & \textbf{+ Formal \texttt{.el} Invariant Exclusions (2,756 pts)} & \textbf{100.0\%} & \textbf{100.0\%} & --- & \textbf{100.0\%} & \textbf{100.0\%} \\
\bottomrule
\end{tabular}%
}
\end{table}

\subsection{Cross-Benchmark 4-Way Ablation Study Results}

\textbf{Table~\ref{tab:ablation}} presents the unified 4-way quantitative ablation study across all five hardware benchmarks (for \texttt{Layout-MultiPipe-Top}, $13{,}879$ is the post-regression residual hole count entering the signoff sweep, URG \texttt{Cond} counts sub-expressions, and \texttt{Eff.\ Closure} is the URG multi-metric \texttt{SCORE} including FSM).

\begin{table*}[!t]
\centering
\caption{Unified 4-Way Quantitative Ablation Study Across 3 OpenTitan IPs and 2 Industrial TLC Subsystems (376,127 Total Points)}
\label{tab:ablation}
\scriptsize
\setlength{\tabcolsep}{3.5pt}
\renewcommand{\arraystretch}{0.93}
\resizebox{\textwidth}{!}{%
\begin{tabular}{llccccccc}
\toprule
\textbf{Benchmark Subsystem} & \textbf{Ablation Variant} & \textbf{Triage Queries / Units} & \textbf{Est.\ LLM Tokens} & \textbf{Line (Raw/Eff)} & \textbf{Branch (Raw/Eff)} & \textbf{Cond/Expr (Raw/Eff)} & \textbf{Toggle (Raw/Eff)} & \textbf{Eff.\ Closure} \\
\midrule
\multirow{4}{*}{\shortstack[l]{\textbf{OpenTitan \texttt{uart\_core}}\\\textit{(7 Modules, 1,982 Pts)}}}
 & Baseline A: Flat LLM Loop & 1,304 raw holes & 527,200 (1.0$\times$) & 73.68\% / 73.68\% & 80.68\% / 80.68\% & 91.55\% / 91.55\% & 38.84\% / 38.84\% & 43.59\% \\
 & Variant B: Macro Only (No Spec) & 37 macro + 7 micro & 34,200 (15.4$\times\downarrow$) & 98.25\% / 98.25\% & 89.77\% / 89.77\% & 94.37\% / 94.37\% & 60.87\% / 60.87\% & 64.43\% \\
 & Variant C: Spec + Micro ($\tau=\infty$) & 1,304 micro AST checks & 418,500 (1.26$\times\downarrow$) & 98.25\% / 98.25\% & 97.73\% / 97.73\% & 97.18\% / 97.18\% & 84.20\% / 84.20\% & 85.82\% \\
 & \textbf{Variant D: Full ARGUS (Ours)} & \textbf{37 macro $\to$ 16 micro $\to$ 70 \texttt{.el}} & \textbf{19,400 (27.2$\times\downarrow$)} & \textbf{98.25\% / 100.0\%} & \textbf{100.0\% / 100.0\%} & \textbf{98.59\% / 100.0\%} & \textbf{96.15\% / 100.0\%} & \textbf{100.00\%} \\
\midrule
\multirow{4}{*}{\shortstack[l]{\textbf{OpenTitan \texttt{i2c\_core}}\\\textit{(6 Modules, 3,906 Pts)}}}
 & Baseline A: Flat LLM Loop & 2,750 raw holes & 1,100,000 (1.0$\times$) & 70.04\% / 70.04\% & 59.77\% / 59.77\% & 74.90\% / 74.90\% & 36.10\% / 36.10\% & 42.81\% \\
 & Variant B: Macro Only (No Spec) & 31 macro + 2 micro & 46,800 (23.5$\times\downarrow$) & 90.69\% / 90.69\% & 82.18\% / 82.18\% & 87.04\% / 87.04\% & 52.94\% / 52.94\% & 60.09\% \\
 & Variant C: Spec + Micro ($\tau=\infty$) & 2,750 micro AST checks & 862,400 (1.28$\times\downarrow$) & 95.55\% / 95.55\% & 91.09\% / 91.09\% & 91.90\% / 91.90\% & 88.45\% / 88.45\% & 89.66\% \\
 & \textbf{Variant D: Full ARGUS (Ours)} & \textbf{31 macro $\to$ 10 micro $\to$ 246 \texttt{.el}} & \textbf{30,200 (36.4$\times\downarrow$)} & \textbf{96.76\% / 100.0\%} & \textbf{92.53\% / 100.0\%} & \textbf{93.52\% / 100.0\%} & \textbf{93.60\% / 100.0\%} & \textbf{100.00\%} \\
\midrule
\multirow{4}{*}{\shortstack[l]{\textbf{OpenTitan \texttt{aes\_cipher\_core}}\\\textit{(14 Modules, 19,147 Pts)}}}
 & Baseline A: Flat LLM Loop & 7,064 raw holes & 2,825,600 (1.0$\times$) & 63.39\% / 63.39\% & 36.19\% / 36.19\% & 56.98\% / 56.98\% & 63.01\% / 63.01\% & 62.66\% \\
 & Variant B: Macro Only (No Spec) & 56 macro + 31 micro & 64,500 (43.8$\times\downarrow$) & 84.82\% / 84.82\% & 77.14\% / 77.14\% & 78.21\% / 78.21\% & 67.81\% / 67.81\% & 68.10\% \\
 & Variant C: Spec + Micro ($\tau=\infty$) & 7,064 micro AST checks & 2,189,800 (1.29$\times\downarrow$) & 94.64\% / 94.64\% & 86.19\% / 86.19\% & 83.80\% / 83.80\% & 77.90\% / 77.90\% & 78.14\% \\
 & \textbf{Variant D: Full ARGUS (Ours)} & \textbf{56 macro $\to$ 20 micro $\to$ 4,120 \texttt{.el}} & \textbf{43,600 (64.8$\times\downarrow$)} & \textbf{94.64\% / 100.0\%} & \textbf{87.14\% / 100.0\%} & \textbf{84.36\% / 100.0\%} & \textbf{78.23\% / 100.0\%} & \textbf{100.00\%} \\
\midrule
\multirow{4}{*}{\shortstack[l]{\textbf{Industrial \texttt{Layout-3D-Subsys}}\\\textit{(16 Modules, 7,970 Pts)}}}
 & Baseline A: Flat LLM Loop & 4,227 raw holes & 1,452,000 (1.0$\times$) & 79.87\% / 79.87\% & 73.67\% / 73.67\% & --- & 56.50\% / 56.50\% & 58.23\% \\
 & Variant B: Macro Only (No Spec) & 192 macro + 4 micro & 68,400 (21.2$\times\downarrow$) & 92.01\% / 92.01\% & 87.50\% / 87.50\% & --- & 60.54\% / 60.54\% & 63.05\% \\
 & Variant C: Spec + Micro ($\tau=\infty$) & 4,227 micro AST checks & 1,120,000 (1.30$\times\downarrow$) & 96.49\% / 96.49\% & 93.88\% / 93.88\% & --- & 62.48\% / 62.48\% & 65.30\% \\
 & \textbf{Variant D: Full ARGUS (Ours)} & \textbf{192 macro $\to$ 14 micro $\to$ 2,756 \texttt{.el}} & \textbf{42,800 (33.9$\times\downarrow$)} & \textbf{98.08\% / 100.0\%} & \textbf{95.21\% / 100.0\%} & --- & \textbf{62.48\% / 100.0\%} & \textbf{100.00\%} \\
\midrule
\multirow{4}{*}{\shortstack[l]{\textbf{Industrial \texttt{Layout-MultiPipe-Top}}\\\textit{(10 Modules, 343,122 Pts)}}}
 & Baseline A: Flat LLM Loop & 13,879 raw holes & 5,551,600 (1.0$\times$) & 90.89\% / 90.89\% & 83.21\% / 83.21\% & 75.34\% / 75.34\% & 91.87\% / 91.87\% & 80.54\% \\
 & Variant B: Macro Only (No Spec) & 42 macro-clusters & 84,000 (66.1$\times\downarrow$) & 92.40\% / 92.40\% & 86.15\% / 86.15\% & 78.10\% / 78.10\% & 92.60\% / 92.60\% & 83.20\% \\
 & Variant C: Spec + Micro ($\tau=\infty$) & 13,879 micro AST checks & 4,163,700 (1.33$\times\downarrow$) & 99.10\% / 99.10\% & 98.67\% / 99.20\% & 86.29\% / 98.80\% & 96.53\% / 99.50\% & 99.15\% \\
 & \textbf{Variant D: Full ARGUS (Ours)} & \textbf{42 macro $\to$ 8 classes $\to$ 13,879 \texttt{.el}} & \textbf{58,000 (95.7$\times\downarrow$)} & \textbf{99.45\% / 100.0\%} & \textbf{98.67\% / 100.0\%} & \textbf{86.29\% / 100.0\%} & \textbf{96.53\% / 100.0\%} & \textbf{100.00\%} \\
\bottomrule
\end{tabular}%
}
\end{table*}

\begin{enumerate}
    \item \textbf{Monotonic Token Compression Scaling (27.2$\times \to$ 36.4$\times \to$ 64.8$\times \to$ 95.7$\times$):} Across the three open-source OpenTitan subsystems alone, as block scale increases from 1,982 points (\texttt{uart\_core}) to 3,906 points (\texttt{i2c\_core}) and 19,147 points (\texttt{aes\_cipher\_core}), ARGUS's $\tau=10$ Macro-Clustering compresses initial macro-cluster query counts by \textbf{35.2$\times$} (\texttt{uart\_core}, $1{,}304 \to 37$), \textbf{88.7$\times$} (\texttt{i2c\_core}, $2{,}750 \to 31$), and \textbf{126.1$\times$} (\texttt{aes\_cipher\_core}, $7{,}064 \to 56$), reducing LLM token consumption by \textbf{27.2$\times$}, \textbf{36.4$\times$}, and \textbf{64.8$\times$} respectively---and reaching \textbf{95.7$\times$} on the 343,122-point \texttt{Layout-MultiPipe-Top} hierarchy.
    \item \textbf{Why \texttt{SpecExtractor} is Mandatory Across All 5 Subsystems (Variant B vs.\ Variant D):} Without \texttt{SpecExtractor} (Variant B), stimulus generation stalls on protocol-coupled and fault-hardened logic: (i) on OpenTitan \texttt{aes\_cipher\_core}, \texttt{aes\_sel\_buf\_chk} triggers a fatal alert (\texttt{alert\_o = 1}) whenever any control signal deviates from the 3-bit sparse encoding (\texttt{SP2V\_HIGH = 3'b011}, \texttt{SP2V\_LOW = 3'b100}); (ii) on OpenTitan \texttt{i2c\_core}, transitioning into \texttt{StretchAcqFull} (\texttt{i2c\_fsm.sv:1201}) requires driving 64 back-to-back target transactions without reading \texttt{acqdata} while maintaining exact SCL/SDA timing (\texttt{thigh}, \texttt{tlow}, \texttt{tsu\_sta}, \texttt{thd\_sta}); and (iii) on \texttt{Layout-3D-Subsys}, Variant B generates $8/31$ scoreboard-failing tests that violate $C_{\max} \le C_{\text{pack}}$ pooling invariants.
\end{enumerate}

\subsection{Formal Exclusion Soundness Verification (21,071 Excluded Points)}

We audited 100\% of the synthesized \texttt{.el} exclusions across all five benchmarks (\textbf{Table~\ref{tab:soundness}}). Across all \textbf{21,071 excluded items}, \textbf{zero reachable coverage points were falsely excluded (100.0\% precision and soundness)}.

\begin{table*}[!t]
\centering
\caption{Taxonomic \& Formal Soundness Audit of Synthesized \texttt{.el} Exclusions Across All Five Benchmarks}
\label{tab:soundness}
\scriptsize
\setlength{\tabcolsep}{4pt}
\renewcommand{\arraystretch}{0.93}
\resizebox{\textwidth}{!}{%
\begin{tabular}{lp{3.6cm}cp{2.9cm}p{6.2cm}c}
\toprule
\textbf{Benchmark} & \textbf{Invariant Category} & \textbf{Excluded} & \textbf{Metric Breakdown} & \textbf{Formal Reachability Proof / Structural Invariant} & \textbf{False Excl.} \\
\midrule
OpenTitan \texttt{uart\_core} & 1. 2-State Enum, Reset, Tie-Offs \& Timeout MSBs & 70 & 1 Line, 1 Expr, 68 Tgl & \texttt{break\_st\_q} 1b enum (1), \texttt{under\_rst} (1), \texttt{fifo\_ctrl '0} tie-offs (14), shift bounds (6), \texttt{rx\_timeout[23:12]} (48) & 0 (100\% Sound) \\
OpenTitan \texttt{i2c\_core} & 2. Sparse FSM Defaults, Stretch Tie-Offs \& MSBs & 246 & 8 Line, 26 Br, 16 Expr, 196 Tgl & \texttt{tcount\_sel\_e}/\texttt{state\_e} defaults (3), 4$\times$\texttt{under\_rst} (4), \texttt{stop\_tx/acq.d '0} (14), counter MSBs (90), sync guards (135) & 0 (100\% Sound) \\
OpenTitan \texttt{aes\_cipher} & 3. Compile-Time \texttt{SecMasking=0} \& Fault Guards & 4,120 & 6 Line, 27 Br, 28 Expr, 4,059 Tgl & \texttt{gen\_no\_masks} (\texttt{prd\_masking}, \texttt{sb\_in/out\_mask} tied to \texttt{'0}: 4,016), \texttt{SBoxImplDom}/\texttt{AES192} params (11), dual-rail guards (93) & 0 (100\% Sound) \\
Industrial \texttt{Layout-3D} & 4. Loop Bounds, Ping-Pong Wrap \& \texttt{ADDR[15:7]} & 2,756 & 6 Line, 18 Br, 2,732 Tgl & $C_{\text{logical}} < 32$ bound (8), 128-row wrap (8), one-hot dtype (8), container MSB/ZP (1,475), \texttt{ADDR\_WIDTH[15:7]} (1,094), $8n$ bits (163) & 0 (100\% Sound) \\
Industrial \texttt{Layout-MultiPipe} & 5. FSM Mutual Exclusion \& Feature Tie-Offs & 13,879 & 286 Blk, 263 Br, 13,263 Cond, 67 Tgl & \texttt{IDLE}/\texttt{DONE} mutual exclusion (6,690), feature-disable tie-off \& open port (3,516), one-hot \texttt{out\_port\_type} (1,706), DRAM/VecMem split (1,474), misc (493) & 0 (100\% Sound) \\
\midrule
\textbf{TOTAL AUDITED} & \textbf{All 5 Hardware Benchmarks} & \textbf{21,071} & \textbf{307 Line/Blk, 334 Br, 13,308 Cond/Expr, 7,122 Tgl} & \textbf{100.0\% Mathematically / Structurally Unreachable} & \textbf{0 (0.0\%)} \\
\bottomrule
\end{tabular}%
}
\end{table*}

\subsection{Comparison Against LLM-for-DV \& Commercial Formal Baselines}

\textbf{Table~\ref{tab:baselines}} compares ARGUS against domain-adapted LLM copilots (\textbf{NVIDIA ChipNeMo}~\cite{liu2024chipnemo}), LLM assertion and RTL repair frameworks (\textbf{AutoSVA}~\cite{orenes2023llm,orenes2021autosva} and \textbf{NVIDIA RTLFixer}~\cite{tsai2024rtlfixer}, alongside \textbf{ChatTest}~\cite{wan2026chattest}), unclustered LLM stimulus generators (\textbf{LLM4DV}~\cite{zhang2023llm4dv}), and commercial formal coverage reachability signoff tools (\textbf{Synopsys VC Formal FCA / Cadence JasperGold Cov}~\cite{synopsys_fca,cadence_jg}). Unlike flat LLM stimulus loops that query uncovered lines independently and violate sparse handshake or temporal invariants, ARGUS bounds every synthesized UVM sequence using \texttt{BlockContract} constraints and enforces a \textbf{3-step progressive sanity gate} prior to regression merge: (1) compile-only syntax and elaboration verification (\texttt{-c}), (2) single-test smoke execution with zero UVM errors/fatals (\texttt{-cr -{}-repeat\_test 1}), and (3) single-VDB delta triage confirming actual target hole closure before cumulative \texttt{.vdb} union merging.

\begin{table*}[!t]
\centering
\caption{Architectural \& Quantitative Comparison Against LLM-for-DV and Commercial Formal Baselines}
\label{tab:baselines}
\scriptsize
\setlength{\tabcolsep}{3pt}
\renewcommand{\arraystretch}{0.90}
\resizebox{\textwidth}{!}{%
\begin{tabular}{>{\raggedright\arraybackslash}p{3.0cm}>{\raggedright\arraybackslash}p{3.4cm}>{\raggedright\arraybackslash}p{2.9cm}>{\raggedright\arraybackslash}p{3.8cm}>{\raggedright\arraybackslash}p{3.1cm}>{\raggedright\arraybackslash}p{2.6cm}}
\toprule
\textbf{Verification Framework} & \textbf{Primary Mechanism} & \textbf{Prompt Context ($>$50K LoC)} & \textbf{Unreachable Dead-Logic Handling} & \textbf{Stimulus Validity \& Gating} & \textbf{Industrial Subsystem Runtime} \\
\midrule
\textbf{NVIDIA ChipNeMo}~\cite{liu2024chipnemo} & Domain-adapted LLM + RAG over specs/docs & Interactive Q\&A snippet scope & None (\emph{Manual engineer review}) & N/A (\emph{Copilot script assist}) & \emph{Human-in-the-loop} \\
\addlinespace[1.5pt]
\textbf{AutoSVA}~\cite{orenes2023llm,orenes2021autosva} / \textbf{RTLFixer}~\cite{tsai2024rtlfixer} & LLM SVA synthesis \& compiler-guided RTL repair & Single-module RTL scope (No URG/VDB) & Local module formal properties only & Compiler / local FPV feedback loop & Manual per-module setup \\
\addlinespace[1.5pt]
\textbf{Flat LLM Loop (LLM4DV)}~\cite{zhang2023llm4dv} & Iterative per-bin LLM stimulus prompting & 1.00$\times$ Baseline (0.53M--5.55M tokens) & None in~\cite{zhang2023llm4dv}; ungrounded waivers risk false exclusions & Simulation bin-hit check only (No spec filter) & Stalls on unreachable invariants \\
\addlinespace[1.5pt]
\textbf{VC Formal FCA / Jasper Cov}~\cite{synopsys_fca,cadence_jg} & Exhaustive bounded / unbounded model checking & N/A (\emph{State-space COI graph}) & \emph{Exact in bounded COI; inconclusive on deep datapaths/SRAMs} & N/A (\emph{Reachability proof only}) & \emph{State-space explosion on large hierarchies} \\
\midrule
\textbf{ARGUS (Proposed)} & \textbf{\texttt{BlockContract} + $\tau$-Switched Macro/Micro + \texttt{.el} Proofs} & \textbf{27.2$\times$--95.7$\times$ tok $\downarrow$ (19k--58k tokens)} & \textbf{AST + SVA Invariant \texttt{.el} Proofs (0 false waivers)} & \textbf{Contract-constrained + 3-step VCS gate (0 dirty)} & \textbf{0.265 s URG triage + $<$1.2 h E2E closure} \\
\bottomrule
\end{tabular}%
}
\end{table*}

\begin{table*}[!t]
\centering
\caption{Enterprise Scalability \& Autonomous Coverage Closure on \texttt{Industrial-ICN-2626} Subsystems}
\label{tab:enterprise}
\scriptsize
\setlength{\tabcolsep}{3pt}
\renewcommand{\arraystretch}{0.90}
\resizebox{\textwidth}{!}{%
\begin{tabular}{>{\raggedright\arraybackslash}p{3.3cm}cc>{\raggedright\arraybackslash}p{2.7cm}>{\raggedright\arraybackslash}p{2.6cm}>{\raggedright\arraybackslash}p{4.6cm}>{\raggedright\arraybackslash}p{3.0cm}}
\toprule
\textbf{Industrial Hierarchy} & \textbf{Scale} & \textbf{Triage} & \textbf{Context Compression} & \textbf{Pre-ARGUS Baseline} & \textbf{Post-ARGUS Closure (L / Br / FSM / Total)} & \textbf{Proven \texttt{.el} Exclusions} \\
\midrule
\textbf{\texttt{Industrial-ICN-2626}} (Full Subsystem) & 2,626 mod & \textbf{0.265 s} (1 core) & \textbf{$>$180$\times\downarrow$} ($<$200 vs.\ 36k tok) & 85.80\% merged & \emph{\textbf{Triage-only:} 291 macro-clusters isolated; $+$3.8\%--4.5\% dead-code lift identified} & 33 dead/tied-off IP stub \& wrapper modules \\
\addlinespace[1.5pt]
\textbf{\texttt{Industrial-Fetch-Top}} (Fetch \& Dispatch) & 59 mod & \textbf{0.042 s} & \textbf{142.0$\times\downarrow$} ($<$250 digest tok) & Sub-block baseline (85.8\% subsys) & \textbf{100.0\% Line / 100.0\% Branch / 100.0\% FSM / 99.32\% Overall} & 15 read-only RdDMA/AXI modules (0 false) \\
\addlinespace[1.5pt]
\textbf{\texttt{Industrial-DRAM-Arb}} (NoC Arbiter) & 15-port AXI & \textbf{0.018 s} & \textbf{115.0$\times\downarrow$} (0.35 ms surrogate) & 8.98\% \texttt{resp\_sel} / 11.14\% \texttt{addr\_sel} & \textbf{100.0\% Line / 100.0\% Branch / 100.0\% FSM / 99.67\% Overall} & 9 read-only write-return slices (0 false) \\
\bottomrule
\end{tabular}%
}
\end{table*}

\subsection{Enterprise Scalability on 2,626-Module Subsystem, Fetch Unit, \& DRAM Arbiter}

To evaluate ARGUS on massive multi-hundred-thousand-line industrial hierarchies where both flat LLM context windows and commercial formal reachability tools time out, we deployed ARGUS's native fast-path AST and Synopsys URG triage engine (\texttt{urg\_parser.py}) on a production advanced-node AI accelerator Interconnect and Memory Controller subsystem (\texttt{Industrial-ICN-2626}, comprising \textbf{2,626 SystemVerilog module definitions}) and performed end-to-end autonomous closure on its \textbf{Instruction Fetch \& Dispatch Unit (\texttt{Industrial-Fetch-Top}, 59 modules)} and \textbf{15-Port DRAM NoC Arbiter (\texttt{Industrial-DRAM-Arb})} (\textbf{Table~\ref{tab:enterprise}}):
\begin{enumerate}
    \item \textbf{Sub-Second Triage Across 2,626 Modules (0.265 s, $>$180$\times$ Token Reduction):} Streaming raw URG coverage reports for \texttt{Industrial-ICN-2626} previously required 81 iterative agent steps, $\sim$15 minutes of wall-clock latency, and $\sim$36,000 output tokens. ARGUS's fast-path parser aggregates all 2,626 module tables in \textbf{0.265 seconds} and emits a structured \texttt{agent\_digest.json} in under 200 tokens (\textbf{$>$180$\times$ token reduction}).
    \item \textbf{Closing the 59-Module \texttt{Industrial-Fetch-Top} Hierarchy to 99.32\% (100\% Line/Branch/FSM):} ARGUS partitioned the 59 Fetch-Top modules into 33 macro-clusters (e.g., dormant CtrlStream instruction fetch FSMs at 36.4\% and multi-engine EngA/EngB arbitration), 10 micro-surgery corner modules (\texttt{rd\_dma\_credit\_ctrl}, \texttt{rd\_dma\_err\_ctrl}), 15 structurally unreachable modules (14 read-only instruction-fetch RdDMA dispatchers with statically tied-off AXI write channels \texttt{AWVALID=0, WVALID=0} plus a non-exclusive \texttt{ARLOCK=0} slave monitor), and 1 top-level wrapper. Synthesizing grounded UVM virtual sequences (\texttt{ctrl\_stream\_stress}, \texttt{multi\_engine\_arb}, \texttt{rd\_dma\_err\_bp}) closed top-level \texttt{Industrial-Fetch-Top} to \textbf{99.32\% overall coverage} with \textbf{100.00\% Line, 100.00\% Branch, 100.00\% FSM, and 99.76\% Condition coverage}.
    \item \textbf{Forensic Dead-Logic Isolation in Auto-Generated Crossbars (\texttt{Industrial-DRAM-Arb}):} On the 15-port DRAM arbiter, a persistent 8.98\% coverage hole in \texttt{resp\_sel} stalled manual DV closure. ARGUS's AST + SVA invariant engine proved that an upstream interconnect generator had symmetrically instantiated write-response (\texttt{bvalid}/\texttt{bresp}) return multiplexers for 9 architecturally read-only RdDMA masters---paths explicitly forbidden by SVA monitor \texttt{No\_bresp\_to\_ro\_dma\_A}. Combining a 14-master concurrent UVM stress test with formal \texttt{.el} exclusion of the 9 dead write-return slices brought \texttt{Industrial-DRAM-Arb} to \textbf{99.67\% effective closure}.
\end{enumerate}

\section{Conclusion}

We presented \textbf{ARGUS}, an autonomous specification-grounded coverage closure and formal invariant exclusion synthesis engine evaluated across \textbf{three open-source LowRISC OpenTitan IPs} (\texttt{uart\_core}, \texttt{i2c\_core}, \texttt{aes\_cipher\_core}) and \textbf{four industrial advanced-node AI accelerator subsystems} (\texttt{Layout-3D-Subsys}, \texttt{Layout-MultiPipe-Top}, \texttt{Industrial-Fetch-Top}, and the 2,626-module \texttt{Industrial-ICN-2626} hierarchy). By combining automated \texttt{BlockContract} extraction with adaptive $\tau$-switched Macro-Clustering, AST Micro-Surgery, and a 3-step progressive UVM sanity gate, ARGUS eliminates the token explosion and illegal-stimulus pitfalls of flat LLM verification loops while overcoming the state-space timeouts of standalone formal reachability tools. Across \textbf{376,127+ core coverage points and 2,679+ total evaluated SystemVerilog modules}, ARGUS reduces LLM token cost by \textbf{27.2$\times$ to $>$180$\times$}, achieves \textbf{zero dirty merges} via contract-constrained 3-step sanity gating, and reaches \textbf{99.32\%--100.0\% effective coverage closure} with \textbf{zero false exclusions}.

\section*{Acknowledgment}

Gemini was used to assist with drafting, structuring, and language editing across Sections I--VI of this manuscript. All experimental designs, RTL and UVM verification runs, formal exclusion audits, statistical analyses, and technical claims were conducted and verified by the author, who takes full responsibility for the accuracy and integrity of the work.

\balance

\end{document}